# Spontaneous gelation of wheat gluten proteins in a food grade solvent

*Mohsen Dahesh[1,2], Amélie Banc[1], Agnès Duri[2], Marie-Hélène Morel[2], and Laurence Ramos[1]**

[1] Laboratoire Charles Coulomb (L2C), UMR 5221 CNRS-Université de Montpellier, Montpellier, F-France.
[2] UMR IATE, UM2-CIRAD-INRA-SupAgro, 2 place Pierre Viala, 34070 Montpellier, France.

* Laurence.ramos@umontpellier.fr

**Abstract**
Structuring wheat gluten proteins into gels with tunable mechanical properties would provide more versatility for the production of plant protein-rich food products. Gluten, a strongly elastic protein material insoluble in water, is hardly processable. We use a novel fractionation procedure allowing the isolation from gluten of a water/ethanol soluble protein blend, enriched in glutenin polymers at an unprecedented high ratio (50%). We investigate here the viscoelasticity of suspensions of the protein blend in a water/ethanol (50/50 v/v) solvent, and show that, over a wide range of concentrations, they undergo a spontaneous gelation driven by hydrogen bonding. We successfully rationalize our data using percolation models and relate the viscoelasticity of the gels to their fractal dimension measured by scattering techniques. The gluten gels display self-healing properties and their elastic plateaus cover several decades, from 0.01 to 10000 Pa. In particular very soft gels as compared to standard hydrated gluten can be produced.

## 1. Introduction

The worldwide rising consumption of meat is not sustainable and access to a wider range of protein-rich foods derived from plant proteins is desirable. Because of their availability at large scale, grain storage proteins issued from wheat (gluten) and corn (zein) are good candidates in the quest for food products based on plant proteins. These proteins share many similarities. They both contain a high amount of proline, and are insoluble in water but soluble in aqueous alcohol. Gluten proteins differ however from zein and exhibit remarkable elastomeric properties. Those properties, which are ascribed to the ability of gluten proteins to form very long polymers, can be regarded as a real asset in the optics of producing protein-based food-products and biomaterials.

The water-insolubility of gluten proteins has precluded their rational study by standard physicochemical approaches and contrarily to globular proteins their structure and phase behavior remain largely unknown. They display moreover a high polymorphism and an extremely broad polydispersity. As several animal elastomeric proteins (elastin from vertebrates, abductin from arthropods, and dragline and flagelliform silks from spiders), they include repeated sequences rich in proline and glycine. The repetitive domains of gluten proteins are also rich in glutamine, which is prone to hydrogen-bonding (Tatham et al., 2001). Gluten proteins include two classes of polypeptides differing in their propensity to form intermolecular disulphide bonds (Fig. 1A). Gliadin, which accounts for half of the gluten proteins are monomeric species and glutenin, the other half, consists in a concatenation of disulfide bond-stabilized polypeptides whose size can reach several millions of kDa (Wrigley, 1996). So far, the complexity of gluten proteins and the lack of knowledge on their structure are clearly scientific bottlenecks that restrict the use of gluten.

In its water hydrated form, gluten is a viscoelastic cohesive mass comprising typically 1.7 g of water per 1 g of protein with an elastic plateau modulus of typically 1 kPa (Cornec et al., 1994; Létang et al., 1999; Ng & McKinley, 2008; Pruska-Kedzior et al., 2008). Making a food product from this matrix is challenging, as gluten displays too high viscosity and elasticity to be easily shaped. Hence fibrous structures could only be obtained at a high energy cost (high temperature and pressure) by extrusion, according to a processing route difficult to control. In addition the negative effects of harsh treatments on the nutritional properties of the gluten based food begin to be recognized (Abe et al., 1984; Rombouts et al., 2012). We will show however that alternative routes might be achievable to make gluten suitable for food processing in mild conditions. In this optics, our strategy is to take one step back and bring a gluten gel to a sol state, which will subsequently gel in a fully controlled fashion. To reach this goal, we have developed an original fractionation procedure yielding an ethanol/water soluble gluten protein mixture comprising a 1:1 weight ratio of gliadin, the monomeric gluten proteins, and glutenin, the polymeric ones. Because the largest most insoluble glutenin polymers are discarded in our process, homogeneous and stable dispersions are obtained for a large range of protein concentrations up to several hundred of mg/ml (Dahesh et al., 2014). The superior quality of samples prepared using ethanol/water mixture as compared to standard gluten is illustrated in fig. 1B-C: dispersing the proteins in water leads to highly heterogeneous samples comprising a viscoelastic mass in an excess of water, whereas homogeneous dispersions and gels are produced using an ethanol/water mixture solvent, a food-grade solvent.

The paper is organized as follows. We first describe the linear viscoelastic properties of the gluten dispersions in an ethanol/water mixture and show that, above a critical concentration, the dispersions exhibit a spontaneous concentration- and time-dependent gelation. We then demonstrate that the sol-gel transition can be quantitatively rationalized using percolation theories and exhibit the hallmarks of critical gelation conventionally obtained for synthetic polymers, and we successfully relate the sample viscoelasticity to structural measurements. Thanks to biochemical assays, we show that the gelation mainly results from slow hydrogen-bonds formation. Finally, we describe the unique rheological properties of the as-prepared gels.

## 2. Materials and Methods

### 2.1. Materials

Gluten fractionation. Native gluten powder (81.94% protein, dry basis) was courtesy of Tereos-Syral (France). A protein fraction representative of gluten in composition (gliadin/glutenin ratio = 1.1), soluble in ethanol/water (50/50, v/v), was extracted according to a protocol previously published by us (Dahesh et al., 2014). In brief, gluten powder (20g) was placed in a centrifuge bottle (volume 250 mL) with 200 mL of 50 % (v/v) ethanol/water gluten and submitted to a continuous rotating agitation (60 rpm at 20°C) for 19 hours. After a 30 minutes centrifugation at 15000g at 25°C, the clear supernatant (yield ~50%) was recovered and placed at 4°C for 24 hours. This led to a phase separation between a light phase (yield ~25%) including gliadin at an unprecedented level of purity and a dense phase (yield ~ 25%) consisting in a 50/50 blend of gliadin and glutenin. The dense phase is the fraction of interest in this study. This fraction was immediately frozen at -18°C before being freeze-dried and grinded. For this fraction, the amount of free thiol groups (0.648 μmoles/g of protein) and of total thiol equivalent groups (106 μmoles/g of protein) were measured according to Morel et al., 2000.

Sample preparation. Samples were prepared by dispersing the required mass of freeze-dried protein fraction in ethanol/water (50/50 v/v). Chemicals, N-Ethylmaleimide (NEM) or β-mercaptoethanol (BME) were eventually added to the solvent before dispersing the proteins. The ageing time is defined as the time elapsed since the sample has been prepared by dispersing the proteins in the solvent. Homogenization was ensured by a rotary shaking overnight at room temperature. Samples were then stored at 20°C for different ageing times.

## 2.2. Experimental methods

High-performance liquid chromatography. Protein size distribution was measured using size exclusion high-performance liquid chromatography (SE-HPLC) performed on an Alliance system equipped with a TSK G4000 SWXL column. Samples were diluted (at about 1mg/ml) in the elution buffer composed of 0.1 M sodium phosphate buffer at pH 6.8 and 0.1% sodium dodecyl sulfate (SDS). To investigate the role of hydrogen bonds, some tests were also performed using an elution buffer containing, in addition to SDS, urea at a concentration of 8M. Elution of the injected sample (20μl) was performed at 0.7 ml/min and the detection of the different species was recorded at a wavelength of 214 nm. The apparent molecular weight calibration of the column was obtained using a series of protein standards with molecular weight in the range 13 to 2 000 kDa according to Dahesh et al., 2014.

Rheology. Rheological measurements were performed on an ARES strain controlled rheometer. A Couette geometry was used for samples concentrations below 300 mg/ml whereas higher concentrations were probed with a plate-plate geometry with rough surfaces to avoid slipping. Low viscosity silicon oil was used to avoid contact between air and the samples and thus prevent solvent evaporation. The following protocol was systematically used: after loading the sample in the shear cell, the less viscous samples were submitted to a shear rate of 1 $s^{-1}$ during 60 s, before being allowed to rest for 30 minutes. The very viscous samples were only allowed to rest for 30 minutes after loading. Viscoelasticity measurements were run after the 30 min rest. We have checked that those protocols led to reproducible initial states. Frequency sweeps, strain sweep and stress relaxation measurements were performed. The extent of the linear regime was probed by running strain sweeps typically at a frequency of 1 rad/s. Frequency sweeps and stress relaxation were performed in the linear regime, with amplitude of the applied shear strain typically between 1 and 20%. Some additional tests were performed by loading a freshly prepared sample in the Couette cell of the rheometer and probing regularly the viscoelasticity of the sample over a period of 8 days. The results are shown in Supplementary data and demonstrate that the sample loading does not play any role on the gelation process.

Scattering experiments. Details concerning the experimental conditions for the scattering experiments are described in Dahesh et al., 2014. In brief, small-angle scattering experiments were conducted on different instruments using either X-ray or neutron beam. For the very low wave vector range ($2.10^{-3}$ - $10^{-1}$ $nm^{-1}$), data were measured using neutrons on the KWS3 instrument operated by FRMII at the Heinz Maier-Leibnitz Zentrum, Garching, Germany. The ID2 beam line of the European Synchrotron radiation facilities (ESRF) at Grenoble, France, was used for the measurement of the intensity scattered in the intermediate $q$-range (0.04 - 1.5 $nm^{-1}$). Finally, an X-ray laboratory set-up was used to measure data at large wave vectors.

Infrared Spectroscopy. Fourier Transformed InfraRed (FTIR) spectroscopy experiments were performed on an Alpha FTIR Bruker apparatus equipped with the single reflection diamond Attenuated Total Reflection (ATR) module. Spectra were recorded by the co-addition of 24 scans at a resolution of 8 $cm^{-1}$. In order to isolate the protein signal, the background was measured with the solvent used to prepare the samples.

All measurements were performed at room temperature.

## 3. Results and discussion

### 3.1. Ageing time- and concentration-dependent viscoelasticity

We investigate the rheological properties of the stable dispersions of the gluten protein blend in water/ethanol (50/50 v/v). Standard linear viscoelasticity measurements of the frequency-dependence of the storage modulus, *G'*, and the loss modulus, *G''*, are shown in Figure 2A-B for samples with different protein concentrations, *C* (105, 185, 317 and 370 mg/ml), and different ageing times (2, 7 and 10 days). At low ageing time/concentration the terminal relaxation of viscoelastic materials is measured: the loss modulus, *G''*, is larger than the storage modulus *G'*, and $G''\sim\omega$ and $G'\sim\omega^2$, with $\omega$ the frequency. As ageing time and concentration increase, both moduli are found to increase and eventually a cross-over from a viscous fluid (*G''*>*G'*) to an elastic solid (*G'*>*G''*) is measured. Interestingly, all data acquired at different concentrations (in the range 105 to 528 mg/ml) and different ageing times (in the range 1 to 57 days) can be collapsed onto one of the two different master curves once the moduli (resp. the frequency) are rescaled by time- and concentration-dependent prefactors, as shown in Figure 2C-D, where *bG'* and *bG''* are plotted as a function of *aω*. Note that few data sets where, in the whole experimentally accessible frequency range, *G'* and *G''* vary as a powerlaw, could not be discriminately associated with one of the two master curves corresponding to pre-gel and post-gel behaviors. Those data sets correspond to the gel point following the Winter and Chambon criterium (Winter & Chambon, 1986).

The scalings evidenced by the building of the master curves nicely demonstrate the self-similarity of the sample viscoelasticity with concentration and ageing time. The pre-gel and post-gel master curves gathering all experiments are shown in Figure S2 (in SI). The two master curves are strikingly similar to those measured for cured synthetic polymers (Adolf & Martin, 1990). The pre-gel master curve (Fig. 2C) obtained at low concentrations and ageing times displays a terminal behavior at small frequency followed at large frequency by a regime where both moduli vary with the same power law with frequency, $G'\sim G''\sim\omega^\Delta$ with a critical exponent $\Delta=0.8$ (Winter & Chambon, 1986; Winters & Mours, 1997). The same critical power law behavior is measured at large frequencies for the post-gel master curve obtained at larger concentrations and ageing times, and at lower frequency a cross-over toward an elastic regime with a frequency-independent storage modulus, $G'_0$, larger than the loss modulus is measured. Note that stress relaxation measurements in the linear regime provide fully consistent results. As shown in Figure 3A, for short ageing time and for time larger than 0.1 s, the relaxation modulus *G(t)* monotonically decays as a power law with time with an exponent -$\Delta$. We found that the numerical value of the parameter $\Delta$ is equal to the value measured at high frequency in a frequency sweep test (Fig. 2) as expected for a powerlaw gel (Ng & McKinley, 2008; Winter & Chambon, 1986). By contrast for long ageing times, *G(t)* decays in a more complex fashion and reaches a non-zero plateau at long times, $G'_p$, which is the signature of a solid-like behavior, and is in quantitative agreement with the low-frequency plateau measured for the storage modulus in oscillatory measurements ($G'_0$). This is illustrated in Figure 3B for a sample of concentration *C*=317 mg/ml whose elastic plateau has been measured at different ageing times. We find that for all ageing times above gelation the elastic modulus measured by a stress relaxation test, $G'_p$, is equal to that measured by a frequency sweep test, $G'_0$, showing the consistency of our measurements and the reproducibility of our data.

Before gelation, we measure that the amplitude of the complex viscosity $\eta^* = \sqrt{\left(\frac{G'}{\omega}\right)^2 + \left(\frac{G''}{\omega}\right)^2}$ increases with ageing time (fig. 4A). $\eta^*$ is found to decrease as a function of the frequency due to the shear thinning of the viscous suspensions. At short ageing time, a plateau at low frequency, $\eta^*_0$, is eventually reached, allowing to extract a zero-frequency complex viscosity. Master curves for a sample at different ageing times are built by normalizing the complex viscosity and the frequency by parameters that depend on the ageing times. This allows the evaluation of the zero-frequency complex viscosity, which is equivalent to the zero-shear viscosity following the so-called Cox-Merz rule, even when it is not accessed experimentally and only the shear thinning regime at higher frequency is measured (Fig. 4B).

We show in Figure 5A that the zero-frequency complex viscosity increases both with ageing time and with concentration. This viscosity is expected to diverge at the gel point, which is not easily measured in the present study due to the very slow processes that take place over several days. We indeed evaluate, based on the transition from pre-gel to post-gel master curves, that the gel point decreases with concentration from ~ 57 days for $C$=185 mg/ml to ~1 day for $C$=238 mg/ml (Fig. 5B). Thanks to the post-gel master curve, one can evaluate the elastic plateau even when the rheological data do not allow to measure it directly as it occurs at too low frequency to be accessed experimentally. Its evolution with ageing time for different concentrations is reported in Figure 5C. The elastic plateau spans several orders of magnitude and displays a fast initial increase followed by a slower increase at larger time. This two-step kinetics is better emphasized in rescaled data (Fig. 5D), where both the elastic plateau is rescaled by the elastic plateau at long time, $G'_{inf}$, and the ageing time by, $t^*$, the cross-over time between the fast and the slow gelation regimes. We measure that $t^*$ continuously decreases with increasing concentration (inset Fig. 5D). A two-step kinetics, as also previously observed for instance for gelatin or chitosan hydrogels (Normand et al., 2000; Hamdine et al., 2006), reflects an initially fast cross-linking regime which subsequently slow down as further cross-linking or reinforcement of the existent cross-linking is hindered.

## 3.2. Critical gelation and link to the structure

Here, we analyze the rheological data in the framework of critical gelation. For a polymer undergoing a cross-linking reaction, $\varepsilon = |p - p_c|$ is defined as a measure of the distance from the gel point, where $p$ is the extent of the cross-linking reaction, and $p_c$ is the critical extent for which an infinite cluster percolate over the whole sample (Winter & Chambon, 1986; Winters & Mours, 1997, Stauffer et al., 1981). A critical polymer gel occurs at $p_c$, for which $G' \propto G'' \propto \omega^\Delta$. In the post-gel state, both the elastic plateau, $G'_0$, and the characteristic time at which $G'$ and $G''$ cross, $\tau$, are predicted to follow critical variations with the distance from the gel point (Martin et al., 1988, 1989): $G'_0 \propto \varepsilon^z$ and $\tau \propto \varepsilon^{-y}$, with critical exponents $z$ and $y$ such that $\Delta = z/y$. This leads to a power law relation between the two viscoelastic parameters, $G'_0$ and $\tau$: $G'_0 \propto \tau^{-\Delta}$, where an exponent that is the opposite to that of the scaling of the complex modulus at the critical point is expected. We have directly tested this prediction.

In Figure 6A, we demonstrate that, for more than 70 experimental conditions (spanning half a decade of concentration and almost two decades of ageing times) $G'_0$ decreases as a power law with $\tau$ over 8 orders of magnitude yielding $G'_0 \propto \tau^{-m}$, with $m = 0.9 \pm 0.2$ , remarkably close to $\Delta = 0.78 \pm 0.07$, the exponent for the evolution of the complex modulus with the frequency, as expected theoretically. Hence our data present the hallmarks of viscoelasticity near critical gel. Those models, initially developed for polymer gels, have been found to satisfyingly account for the viscoelasticity of a large variety of synthetic polymers, including polydimethylsiloxane (Chambon & Winter, 1987, polybutadienes (Mours & Winter, 1996), or polyacrylamide (Larsen & Furst, 2008) and also of biomacromolecules, including alginate (Matsumoto et al., 1992), chitosan (Hamdine et al., 2006), gelatin (Guo et al., 2003), and peptides (Larsen & Furst, 2008; Cingil et al. 2015).
A key ingredient of the critical gel model is the fractal nature of the gel. We have directly checked this assumption using scattering experiments, combining wide and small-angle X-ray scattering and ultra-slow small-angle neutron scattering. Our data (fig. 6B) show that, over almost two decades, the scattered intensity decays as a power law with the scattering vector, $q$: $I \sim q^{-d_f}$, proving the fractal structure of the gluten gel with a fractal dimension $d_f = 2$. We find the same values for all gels with different concentrations. Note that this value is precisely the one theoretically expected for a branched polymer in good solvent as given by lattice-animal models (Martin et al., 1989). This value is moreover fully consistent with our structural measurements in dilute regime (Dahesh et al., 2014) that demonstrate that the protein extract used here display in an ethanol/water 50/50 solvent the structural and dynamical hallmarks of polymer coils in good solvent conditions. Furthermore, models have been developed to relate the critical exponent of the viscoelasticity to structural properties of the polymer near the gelation point. In the case of unscreened excluded-volume interactions percolation theory predicts $\Delta = d/(d_f + 2)$, where $d=3$ is the space dimension and $d_f$ is the fractal dimension of the polymer (Muthukumar, 1989). With $d_f = 2$, one predicts $\Delta = 0.75$, very close to our experimental values ($\Delta = 0.78$). Note that if we assumed instead screened excluded-volume interactions we would expect $\Delta = [d(d + 2 - 2d_f)]/[2(d + 2 - d_f)] = 0.5$ for $d_f = 2$, hence very different from our experimental findings. This unambiguously confirms that the hydrodynamic interactions between the polymer chains are screened, as expected for polymer in good solvent conditions. Quantitative direct connections between structure and visco-elasticity for near-critical gels are very scarce in the literature (Matsumoto et al., 1992; Takenata et al., 2004). Here, our structural and rheological measurements in a concentrated regime provide consistent results in excellent agreement with structural measurements of the protein suspension in a dilute regime.

### 3.3. Origin of the spontaneous gelation process

To get some insight about the physical origin of the slow gelation, we first have measured over ageing time the molecular weight distribution of the proteins using high-performance liquid chromatography (HPLC) in a SDS-buffer (this buffer is expected to disrupt all weak bonds). Data (Fig. 7) clearly show a weak but continuous decrease of the peak corresponding to low molecular weight (MW) proteins (gliadin, MW around 30 kDa) and the region corresponding to (short) glutenin polymers (MW in the range 80-300 kDa). By contrast, the excluded peaks related to very large assemblies of proteins increases. New interchain disulfide bonds arising from disulfide /thiol exchange or thiol oxidation, as observed upon heating and/or mixing of gluten (Morel et al., 2002; Auvergne et al., 2008), could be at the origin of our observations.

Accordingly, in the presence of a chemical (N-Ethylmaleimide, NEM) that blocks the free thiol groups and, thus, prevents any further disulfide-bonds to occur and hinders completely exchanges between existing disulfide bonds (Fig. 8A), we indeed measure a lower content of high molecular weight species. Moreover, we find that the HPLC profile of a 10 days old sample with NEM superimposes perfectly with the one of a 1-day old sample without NEM (Fig. 8B). Interestingly, we nevertheless find that the building-up of a gel still occurs. Indeed, for a sample containing an excess of NEM (1 mM, which corresponds to roughly 3.7 times more NEM than free thiol groups in the sample) an elastic modulus is measured. Moreover, the modulus is found to continuously increase with ageing time (fig. 8D). These data prove that an increase in disulfide interchain bonding is not essential for the gelation. The slightly lower elastic modulus, as compared to a reference sample without NEM, might nevertheless suggest a minor contribution of disulfide interchain bonding to the sample elasticity. Experiments are also performed with β-mercaptoethanol (BME) which is a chemical commonly used to cleave disulfide bonds (Fig. 8A). BME shifts the thiol/disulfide balance of proteins, through the reduction of their inter- or intra-molecular disulfide bonds. In catalytic amount it can promote the reshuffling of protein disulfide bonds through thiol/disulfide exchange and therefore could assist protein unfolding. When used at a concentration of 3 mM in the solvent, which corresponds to about 13% of the total disulfide bonds, BME induces a shift of molecular weight distribution towards smaller protein species indicating a partial breakdown of inter-molecular disulphide bonds (Fig. 8B) but leads to a faster kinetics for the gel build-up (Fig. 8D), presumably thanks to a faster shuffling of the S-S bonds yielding a facilitated gelation. Notably however, the final elastic plateau is comparable to that of the reference sample, suggesting that BME does not modify the number and strength of the junctions in the polymer gel. By contrast, larger quantities (0.3 M), which corresponds to a 13 fold excess of BME compared to the total number of disulfide bonds, cleave all the disulfide bonds, allowing a totally different type of gel to form very rapidly (Fig. 8D), presumably through the formation of intermolecular β-sheets (Fig. 8C). Indeed, FTIR spectra of gels prepared with the different thiol modifiers show that the amide I of protein gels is nearly unmodified in presence of NEM or 3 mM BME whereas in presence of large quantities of BME the shape of the amide I band is totally different and dominated by a contribution at 1620 cm$^{-1}$ assigned to β-sheets (Fig. 8C).

Conclusive results are derived from experiments using urea, a chemical that destabilizes H-bonds. A gel sample readily dissolves in an ethanol/water solvent containing 8 M urea, whereas it keeps its integrity when put into contact with an ethanol/water solvent (Fig. 9A) demonstrating unambiguously that H-bonds formation is the motor for gelation, as also assumed for standard water hydrated gluten (Ng et al., 2011). Size-exclusion chromatography confirms moreover that the MW size distribution profile is preserved by urea addition (Fig. 9B). The H-bonding could occur between the side amide groups of the glutamine aminoacids which are abundant in the repetitive domains of the gluten proteins (Fig. 1A) and are known to be prone to interact through H-bonds (Rhys et al., 2012). Note however that contrarily to what was observed with BME (0.3M), no modification of the infrared spectra could be detected with aging time. We believe that, in this case, the long polymer chains prevent the building of cooperative H-bonds as in interchain β-sheets but instead a few isolated H-bonds forms, which are responsible to the slow gelation of the samples, but are not sufficient to lead to a significant change in the infrared signal with time.

### 3.4. Rheological properties of the novel gluten gels

We have quantified the rheological properties of the gels after long ageing times, once the mechanical properties do not significantly evolve any more.
We measure that the elastic plateau at long time, $G'_{inf}$, can be varied over a much extended range of numerical values, from $10^{-2}$ Pa to $10^4$ Pa (fig. 10A). As shown the elastic modulus grows very rapidly with concentration as it increases by almost 6 orders of magnitude while the concentration varies by a factor of 2. For concentration above 400 mg/ml, the elastic plateau, $G'_{inf}$, increases smoothly with concentration, whereas below this concentration, the dependence of $G'_{inf}$ with concentration is very strong. Our data suggest that the plateau increases exponentially with the protein concentration. Interestingly, an exponential increase of the elastic modulus with concentration has been recently observed with bicontinuous solid-stabilized emulsions (Witt et al., 2013) and with protein gels formed by spinodal decomposition (Gibaud et al., 2013), where it has been interpreted using a model for a porous medium (Pal, 2005). Given the structure of the gels investigated here, an analogy with a porous medium is not obvious. At the present time the physical origin of the very strong dependence of the elastic modulus with concentration remains to be clarified.
On the other hand, our data can be compared with existing data for standard hydrated gluten. For a protein concentration of 400 mg/ml, the typical value for hydrated gluten, the elastic modulus of our novel gel is of the order of 4000 Pa, yielding a three to five-fold stronger gel than standard hydrated gluten (800-1500 Pa) (Ng & McKinley, 2008; Pruska-Kedzior et al., 2008). Hence, substitution of water by an ethanol/water mixture does not alter the viscoelastic characteristics of gluten proteins. However, in striking contrast with hydrated gluten, we are able to produce gels at unprecedented low gluten protein concentrations thus generating materials encompassing a broad range of lower elastic plateau values as compared to standard gluten.
The novel gluten gels also exhibit a surprisingly large linear regime, as shown in Figure 10B where both the storage and loss modulus are found to be independent of the strain amplitude up to $\gamma_c \approx$ 300%, again in sharp contrast with a wheat dough ($\gamma_c < \sim 1\%$) (Cornec et al., 1994; Hibberd & Wallace, 1966; Phan-Tien et al., 1997; Uthayakumaran et al., 2002) and a standard hydrated gluten ($\gamma_c$ is typically a few %) (Cornec et al., 1994; Létang et al., 1999; Uthayakumaran et al., 2002). Moreover, the gels can be extended considerably before fracturing, pinpointing at a ductile behavior. In addition, they exhibit self-healing, highlighting the reversible/physical links building up the gel (fig. 10C). These features are in sharp contrast with other gels made with large amounts of BME, which are very brittle without self-healing properties, and for which all proteins have been cleaved (fig. 8B). Those contrasting results highlight the crucial role of the native polymeric species in the remarkable rheological properties of the novel gluten gels.

## 4. Conclusions

We have shown that water/ethanol soluble blends of gliadin and glutenin (~ 50/50) extracted from gluten undergo a spontaneous gelation, which is driven by the formation of H-bonds. We have quantitatively rationalized the gelation using theoretical percolation models. The mechanical properties of the as-prepared gels can be tailored over an unprecedented range. They display shear elastic modulus spanning almost six orders of magnitude and exhibit an unexpectedly large linear elastic domain, in striking contrast with conventional hydrated gluten. We expect that those novel gels could serve as starting materials for the processing of plant-based protein food products. In the form of biodegradable glue or as electrospun fibers the protein blend could also find non-food applications. As such, the protein fraction investigated here constitutes a promising model to understand the molecular basis of the gluten elastic behavior. Contrarily to simple gluten/water mixture, which pre-exists as a gel, it offers the possibility to investigate in great details the interaction potential of the two main gluten protein classes, gliadin and glutenin. How screening of protein charges by salt and pH alter the gel formation kinetic and its elastic properties will be the subject of our future investigations.

**Acknowledgements**
We thank J. Bonicel (IATE) for help in the SE-HPLC, P. Dieudonné (L2C) for the in-house SAXS measurements and C. Charbonneau (L2C) for help during the neutron scattering measurements. We acknowledge the European Synchrotron Radiation Facility for provision of synchrotron radiation facilities. We would like to thank Pawel Kwasniewski for assistance in using beamline ID02 at ESRF and Zhendong Fu for assistance during the neutron scattering experiments. This work was supported by the Laboratoire of Excellence NUMEV (ANR-10-LAB-20) and INRA (CEPIA department). Besides, this research project has been supported by the European Commission under the 7th Framework Program through the "Research Infrastructures" action of the Capacities Program, NMI3-II, Grant Agreement number 28388 to perform the neutron scattering measurements at the Heinz Maier-Leibnitz Zentrum (MLZ), Garching, Germany.

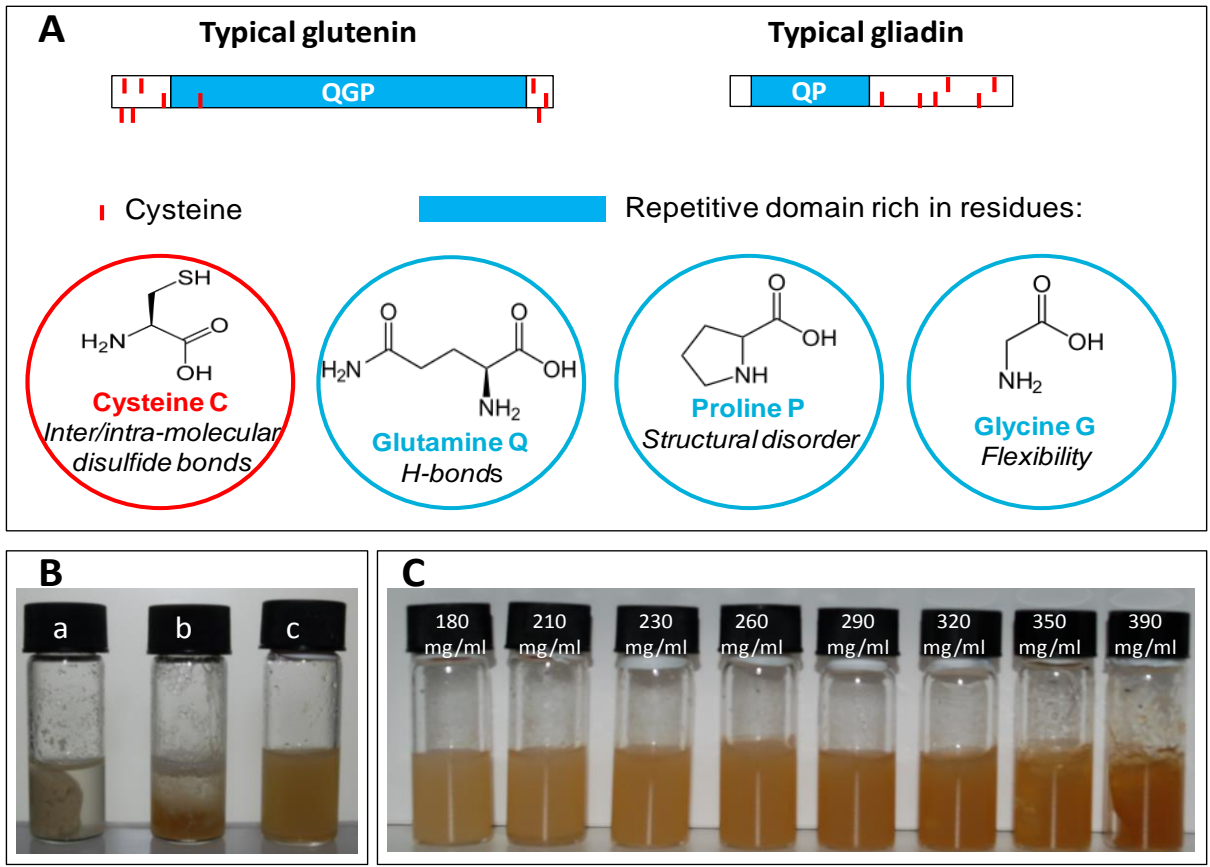

**Figure 1: Gluten proteins and samples**
(A) Typical primary structures of both families of gluten proteins: gliadins and glutenins. Repetitive domains rich in proline (P), glycine (G) and glutamine (Q) residues are highlighted in blue. Cysteine residues involved either in intramolecular (inner lines) or intermolecular (outer lines) disulfides bonds are represented as short vertical rods (in red). Gliadins contain 0, 6 or 8 cysteines involved in intra-molecular disulphide bonds. Glutenins contain 4, 5 or 7 cysteines involved in intramolecular bonds and 1 or 3 cysteines involved in intermolecular disulphide bonds. (B) Comparison of protein dispersions: samples are prepared by dispersing (a) commercial gluten (the starting material used for protein fractionation) in water, (b) the gluten protein extract in water, and (c) the gluten protein extract in a 50/50 (v/v) water/ethanol solvent. In the three samples, the protein concentration is $C$=210 mg/ml. (C) Homogeneous gluten gels, with various protein concentrations as indicated on the picture, produced by dispersing the gluten protein extract in a 50/50 (v/v) water/ethanol solvent.

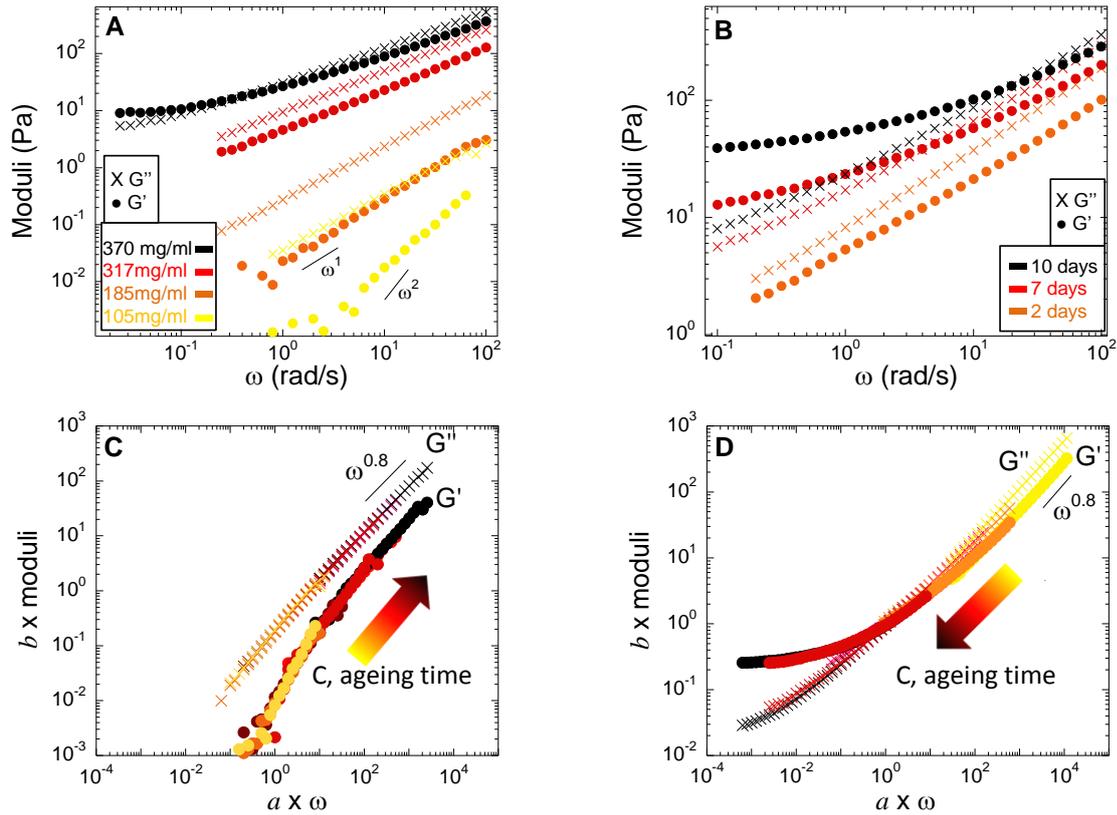

**Figure 2: Viscoelasticity of the gluten suspensions and gels**
Evolution with frequency of the storage ($G'$) and loss ($G''$) moduli for samples with different concentrations and ageing times. Data were acquired using standard shear rheometry in the linear regime. (A) Evolution of $G'$ and $G''$ for 1-day old samples with different concentrations as indicated in the legend. (B) Evolution of $G'$ and $G''$ for a sample with a protein concentration $C=317$mg/ml, at different ageing times as indicated in the legend. (C) Pre-gel master curve. (D) Post-gel master curve. Master curves (C,D) were built shifting data horizontally and vertically using scaling factors $a$ and $b$.

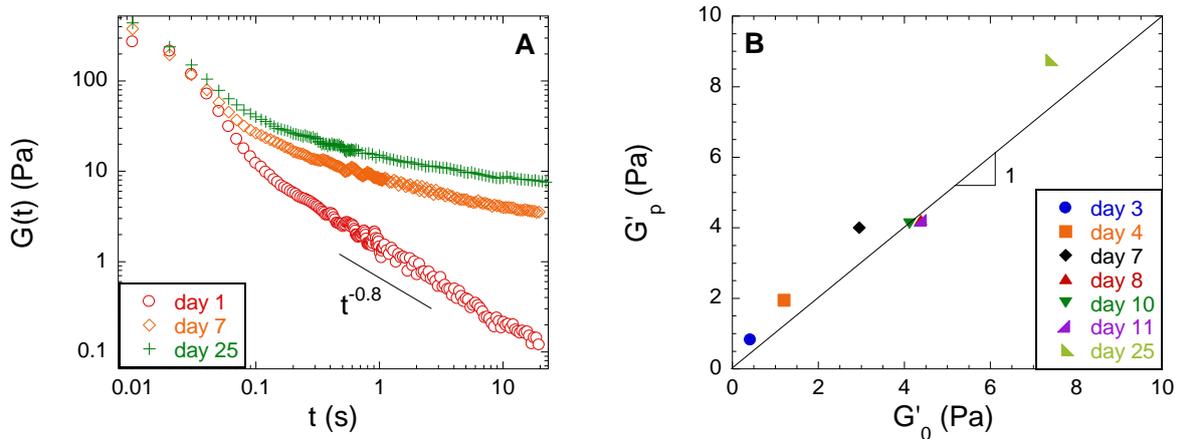

**Figure 3: Stress relaxation of the gluten suspensions and gels**
(A) Stress relaxation in the linear regime of a gel ($C=317$ mg/ml) at different ageing times. The amplitude of the sudden imposed shear strain was well in the linear regime (between 5% and 50%). (B) Comparison of the elastic moduli of samples at different ageing times measured by shear stress relaxation tests ($G'_p$) and by frequency sweep tests ($G'_0$). Symbols are experimental data and the line corresponds to $G'_p = G'_0$.

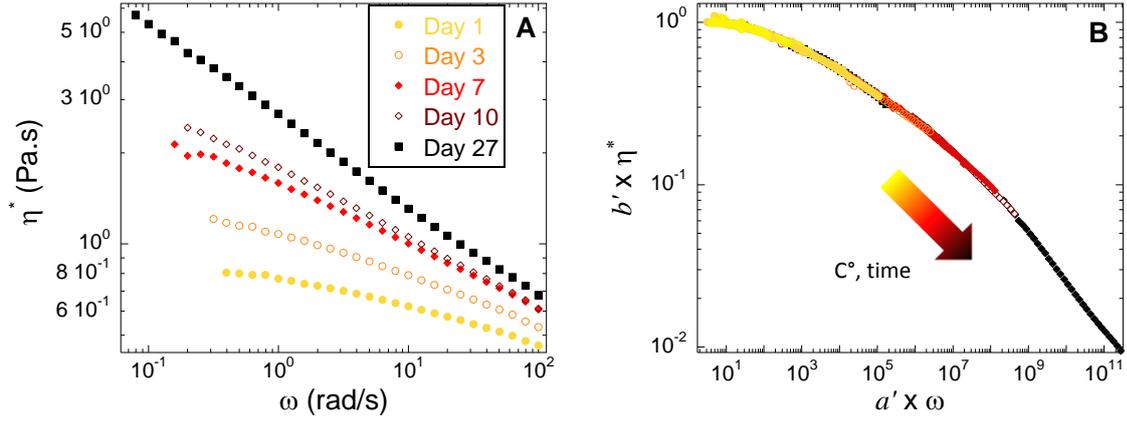

**Figure 4: Complex viscosity of the gluten suspensions**
(A) Complex viscosity $\eta^*$ of samples with concentration $C$=211 mg/ml, for different ageing times (as indicated in the legend) as a function of the frequency. (B) Master curve obtained shifting data horizontally and vertically using scaling factors $a'$ and $b'$. The zero shear complex viscosity is $\eta_0^*=1/b'$.

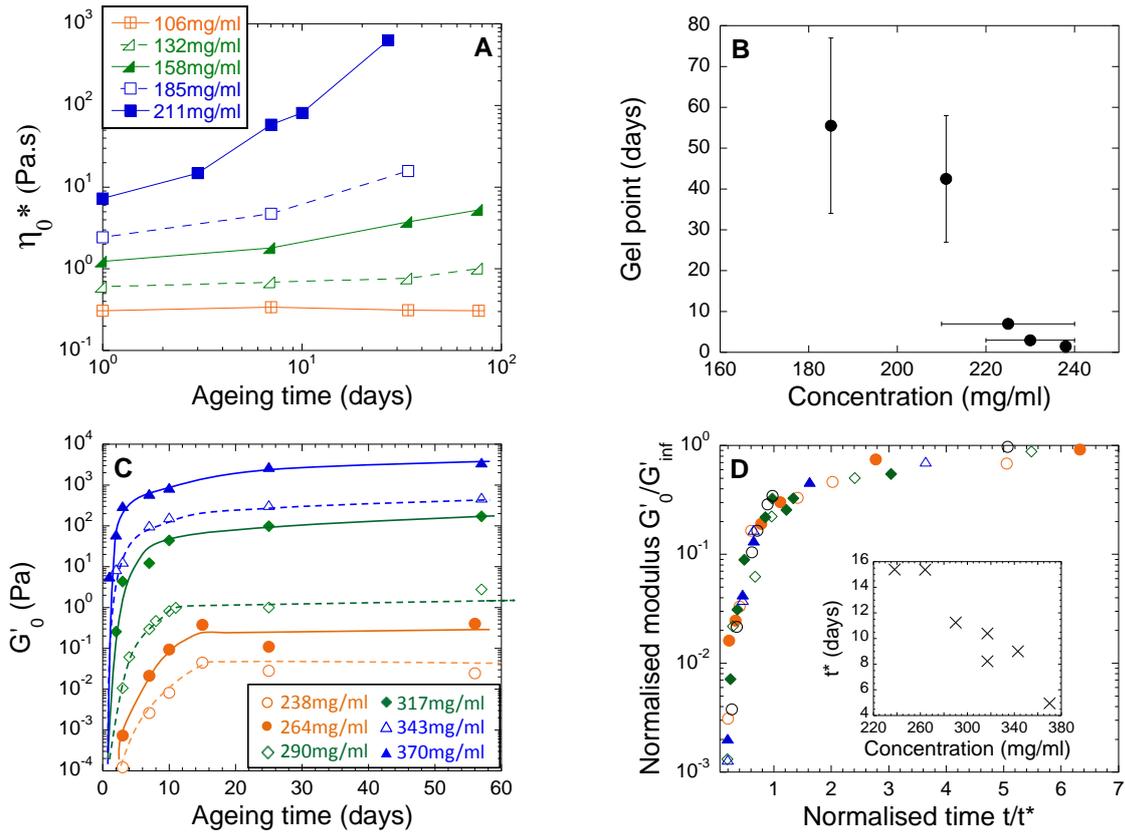

**Figure 5: Kinetics and gel point**
(A) Zero shear complex viscosity $\eta_0^*$ of pre-gel samples of different concentrations as indicated in the legend as a function of ageing time. (B) Evolution of the gel point as a function of protein concentration. (C) Low frequency elastic modulus $G'_0$ of post-gel samples of different concentrations as indicated in the legend as a function of ageing time. Lines are guides for the eyes. (D) Master curve of post-gel kinetics, obtained from rescaling the data shown in (C). The insert displays the evolution of the characteristic time $t^*$ used to produce the main plot.

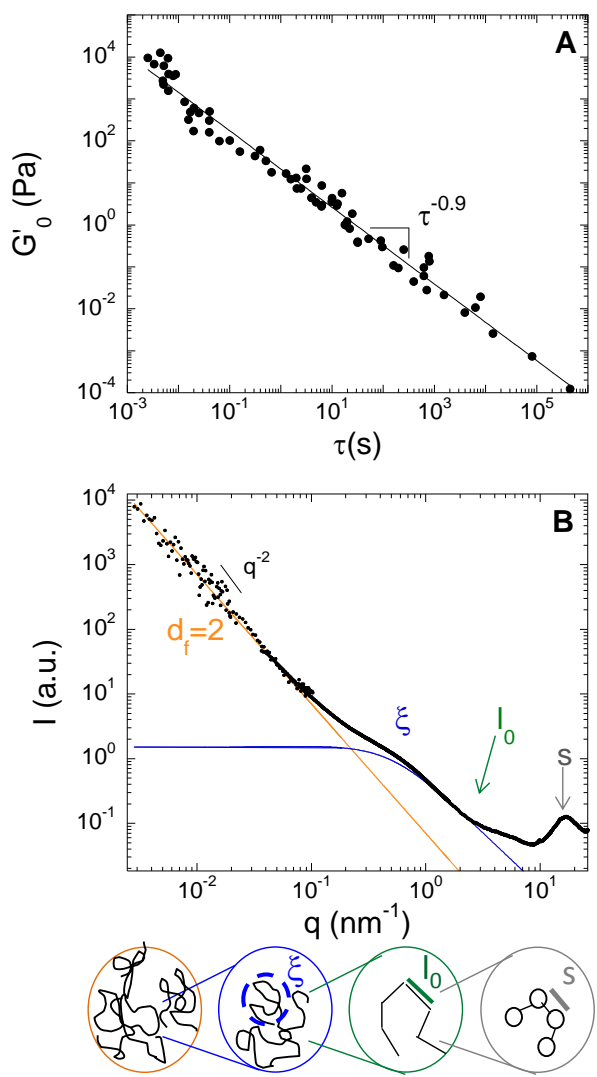

**Figure 6: Critical gelation and link to the structure of the gel**
(A) Low frequency elastic modulus of gels as function of the relaxation time at the crossover between *G'* and *G''*. Symbols correspond to data points evaluated from the scaling factors, *a* and *b*, used to fall the frequency-dependent complex moduli onto the post-gel master-curve. The post-gel master curve gathering all experimental data is shown in SI (Fig. S1). The power law evolution (black line) is characteristic of near-critical gel visco-elasticity (B) Scattering profile of a sample (*C*=290 mg/ml) as obtained from combining several scattering techniques. The profile is typical of a polymer gel in good solvent conditions and evidences a hierarchical structure (as schematized in the cartoon). At large scattering vectors the wide peak corresponds to the liquid like order of the sample, the change of slope at q~3nm$^{-1}$ gives the persistence length of polymeric chains (l$_0$=0.7 nm) and the transition around q~5nm$^{-1}$ is characteristic of the polymer blob size ($\xi$=1.8 nm). The power law decay at small wave vector indicates large scale heterogeneities characterized by a fractal dimension $d_f$=2. More details are provided in Dahesh et al., 2014.

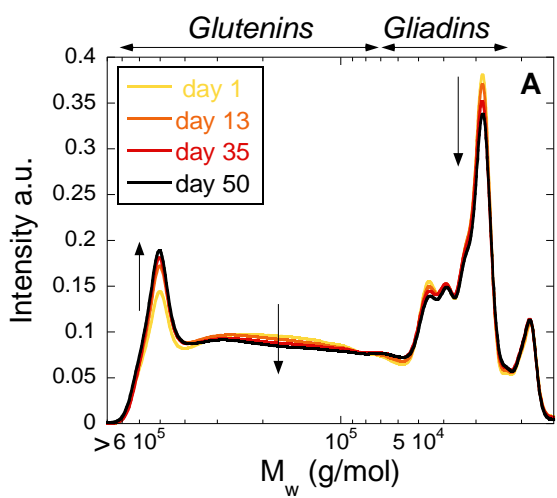

**Figure 7: Ageing of the gluten gels**
High-performance liquid chromatography (HPLC) profiles of samples at different ageing times. Arrows indicate the evolutions with ageing time.

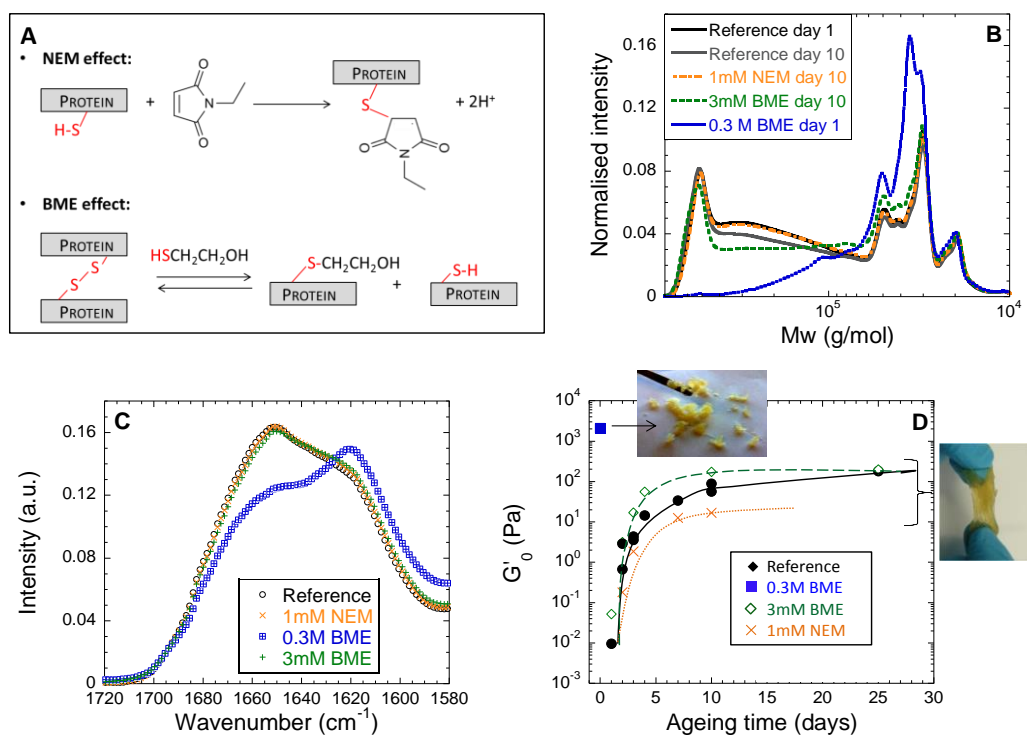

**Figure 8: Role of di-sulfide bonds and thiol modifiers on the gluten gels**
(A) Chemical reactions of N-Ethylmaleimide (NEM) and β-mercaptoethanol (BME) with proteins containing free thiols and disulfide bond respectively. (B) HPLC profiles of fresh and 10 days old samples. Reference samples correspond to pristine samples (protein in water/ethanol) without additional chemicals. The other samples include a fresh sample containing 0.3 M BME and 10-days old samples containing 1 mM NEM or 3mM BME. Data are normalized by the protein concentration in the gel. (C) Amide I band of 10 days old samples aged with NEM or BME measured by infrared spectroscopy. (D) Comparison of the time-evolution of the low frequency elastic modulus of samples with concentration $C$=317mg/ml prepared without thiol modifiers (reference sample) and with different thiol modifiers: 0.3M of BME, 3mM of BME and 1 mM of NEM. The pictures illustrate the brittle structure of the gel prepared with 0.3M of BME compared to the ductility of the gels obtained in the other cases.

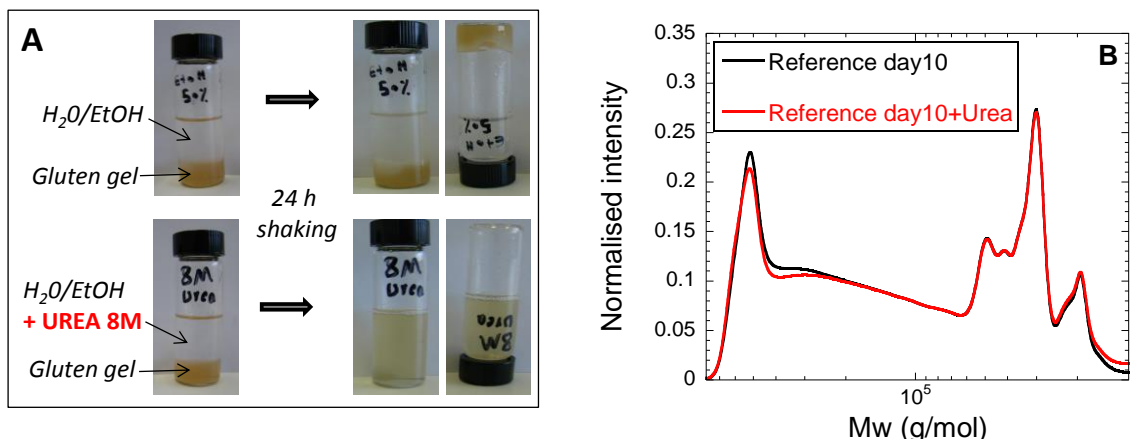

**Figure 9: Impact of urea on the gluten gels**
(A) Picture of a 10 days old gel prepared with a protein concentration $C$=317mg/ml, mixed with an excess of solvent on the top, and with an excess of solvent containing 8M of urea on the bottom. (B) High-performance liquid chromatography (HPLC) profiles of a 10-days old gluten gel dispersed in SDS buffer (reference sample) or in a 8M urea buffer.

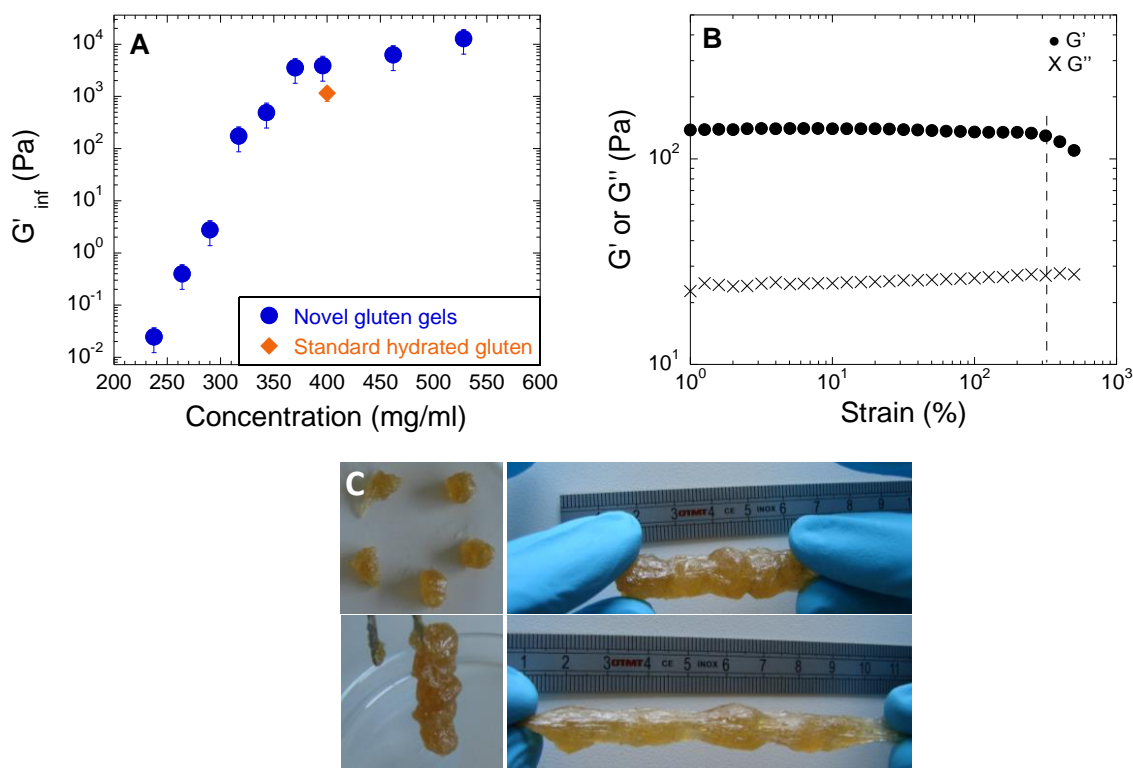

**Figure 10: Mechanical characteristics of the gluten gels**
(A) Low frequency elastic modulus of gels obtained at an infinite ageing time, $G'_{inf}$, as a function of protein concentration. The diamond displays the characteristic value obtained for standard hydrated gluten. (B) Strain sweep of a 56-day old gel of concentration $C$=310 mg/ml illustrating the extremely large linear regime of the gels which can reach more than 300%. (C) Pictures illustrating the self-healing behavior of gels (here an 8-day old gel with $C$=537 mg/ml). Top left: pieces of gel that are used to form the healed material (bottom left). Right: stretching of the healed sample after a 10 minutes rest.

# Supplementary data

**Comparison between in-situ ageing and ex-situ ageing**

Because of the long ageing times, samples usually age in a hermetically closed container (ex-situ gelation). Aliquots of those samples are regularly loaded in the rheometer to probe their viscoelasticity, following a protocol described below in the Method section. To check for the possible role of the sample loading on the aging behavior, some additional tests were performed by loading a freshly prepared sample in the Couette cell of the rheometer and probing regularly the viscoelasticity of the sample over a period of 8 days.

We quantitatively compare the viscoelasticity for a sample allowed to age in-situ to that of a sample that has aged ex-situ. We show in figure S1 comparable evolution with ageing time of the zero-frequency elastic plateaus for in-situ and ex-situ gelation of a sample with concentration $C= 317$ mg/ml. This proves that our protocol leads to results comparable to those obtained when a sample is allowed to age directly in the rheometer cell with a layer of silicon oil on the top to prevent evaporation of the solvent (in-situ gelation).

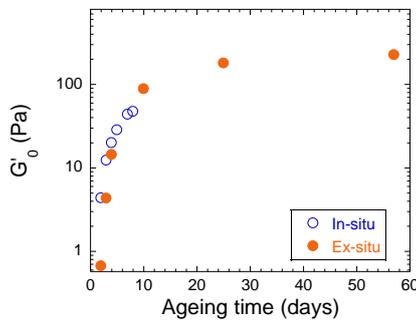

**Figure S1:** Comparison of $G_0'$ measured for a sample with concentration $C=317$ mg/ml at different ageing times using different protocols. The sample has aged either in a flask and has been loaded in the rheological cell before each measurement (ex-situ), or has aged directly in the Couette cell of the rheometer with a silicon oil drop on the top to prevent solvent evaporation (in-situ).

**Pre-gel and post-gel master curves**

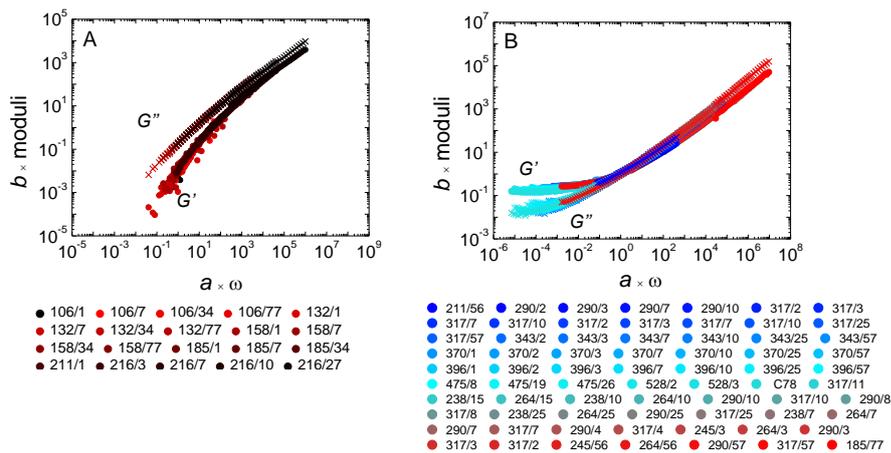

**Figure S2:** (A) Pre-gel mastercurve plotted with data obtained with 20 different measurements. The concentration range of samples was (105-211) mg/ml and their ageing time was in the range (1-77) days. (B) Post-gel mastercurve plotted with data of 69 measurements. The concentration range of samples was (185-528) mg/ml and their ageing time was in the range (1-77) days. In A and B, the numbers x/y in the caption indicate the concentration in mg/ml (x) and the ageing time in days (y).